\documentclass[twocolumn]{aastex631}

\usepackage{multirow}

\newcommand{\thisfrb}{FRB20190520B}
\newcommand{\frbhg}{HG20190520B}

\newcommand{\fobs}{\ensuremath{f_\mathrm{obs}}}

\newcommand{\hMpc}{\ensuremath{h^{-1}\,\mathrm{Mpc}}}
\newcommand{\kpc}{\ensuremath{\mathrm{pkpc}}}

\newcommand{\zfrb}{\ensuremath{z_\mathrm{frb}}}
\newcommand{\zgrp}{\ensuremath{z_\mathrm{grp}}}
\newcommand{\rvir}{\ensuremath{r_{200}}}
\newcommand{\rmax}{\ensuremath{r_{max}}}

\newcommand{\effgee}{\ensuremath{\widetilde{F}G}}
\newcommand{\eff}{\ensuremath{\widetilde{F}}}

\newcommand{\modelone}{\ensuremath{\rmax = \rvir}}
\newcommand{\modeltwo}{\ensuremath{\rmax = 2\,\rvir}}
\newcommand{\mstar}{\ensuremath{M_\star}}
\newcommand{\mhalo}{\ensuremath{M_\mathrm{halo}}}
\newcommand{\msun}{\ensuremath{M_\odot}}
\newcommand{\lgmh}{\ensuremath{\log_{10}(M_\mathrm{halo}/M_\odot) }}

\newcommand{\lgmhnu}{\ensuremath{\log_{10}(M_\mathrm{halo}) }}

\newcommand{\ms}{\ensuremath{\mathrm{ms}}}

\newcommand{\dll}{d_\mathrm{LL,\perp}}
\newcommand{\dpar}{d_\mathrm{LL,\parallel}}

\newcommand{\dmunits}{\ensuremath{\rm pc \, cm^{-3}}}

\newcommand{\dmhalo}{\ensuremath{\mathrm{DM}_\mathrm{halo}}}
\newcommand{\dmhalos}{\ensuremath{\mathrm{DM}_\mathrm{halos}}}

\newcommand{\dmaigm}{\ensuremath{\langle {\rm DM}_{\rm IGM} \rangle}}

\newcommand{\dmfrb}{\ensuremath{\mathrm{DM}_\mathrm{FRB}}}
\newcommand{\dmigm}{\ensuremath{\mathrm{DM}_\mathrm{IGM}}}

\newcommand{\dmhost}{\ensuremath{\mathrm{DM}_\mathrm{host}}}

\newcommand{\dmhostcgm}{\ensuremath{\mathrm{DM}^\mathrm{cgm}_\mathrm{host}}}
\newcommand{\dmhostin}{\ensuremath{\mathrm{DM}^\mathrm{in}_\mathrm{host}}}
\newcommand{\dmmw}{\ensuremath{{\rm DM}_{\rm MW}}}

\newcommand{\fhot}{\ensuremath{f_{hot}}}
\newcommand{\figm}{\ensuremath{f_{igm}}}

\newcommand{\nrich}{\ensuremath{N_\mathrm{rich}}}
\newcommand{\nrichtrue}{\ensuremath{N^\mathrm{true}_\mathrm{rich}}}

\newcommand{\nrichmock}{\ensuremath{N^\mathrm{mock}_\mathrm{rich}}}

\submitjournal{ApJL}

\shorttitle{FRB190520 Foreground Clusters}
\shortauthors{Lee et al.}
\graphicspath{{./}{figures/}}
\usepackage{amsmath}
\usepackage{appendix}
\usepackage{tablefootnote}

\newcommand{\dered}{\mathrm{dered}}

\begin{document}

\title{The FRB20190520B Sightline Intersects Foreground Galaxy Clusters}

\correspondingauthor{Khee-Gan Lee}
\email{kglee@ipmu.jp}

\author[0000-0001-9299-5719]{Khee-Gan Lee}
\affiliation{Kavli IPMU (WPI), UTIAS, The University of Tokyo, Kashiwa, Chiba 277-8583, Japan}
\affiliation{Center for Data-Driven Discovery, Kavli IPMU (WPI), UTIAS, The University of Tokyo, Kashiwa, Chiba 277-8583, Japan}

\author[0000-0003-0574-7421]{Ilya S. Khrykin} 
\affiliation{Kavli IPMU (WPI), UTIAS, The University of Tokyo, Kashiwa, Chiba 277-8583, Japan}
\affiliation{Center for Data-Driven Discovery, Kavli IPMU (WPI), UTIAS, The University of Tokyo, Kashiwa, Chiba 277-8583, Japan}

\author[0000-0003-3801-1496]{Sunil Simha}
\affil{University of California - Santa Cruz
1156 High St.
Santa Cruz, CA, USA 95064}

\author[0000-0002-5934-9018]{Metin Ata}
\affiliation{
The Oskar Klein Centre, Department of Physics, Stockholm University,  \\ AlbaNova University Centre, SE 106 91 Stockholm, Sweden
}

\author[0000-0002-0298-8898]{Yuxin Huang}
\affiliation{Kavli IPMU (WPI), UTIAS, The University of Tokyo, Kashiwa, Chiba 277-8583, Japan}

\author[0000-0002-7738-6875]{J. Xavier Prochaska}
\affil{University of California - Santa Cruz
1156 High St.
Santa Cruz, CA, USA 95064}
\affil{Kavli IPMU (WPI), UTIAS, The University of Tokyo, Kashiwa, Chiba 277-8583, Japan}
\affil{Division of Science, National Astronomical Observatory of Japan,
2-21-1 Osawa, Mitaka, Tokyo 181-8588, Japan}

\author[0000-0002-1883-4252]{Nicolas Tejos}
\affil{Instituto de F\'isica, Pontificia Universidad Cat\'olica de Valpara\'iso, Casilla 4059, Valpara\'iso, Chile}

\author[0000-0001-5703-2108]{Jeff Cooke}
\affil{Centre for Astrophysics and Supercomputing, Swinburne University of Technology, Hawthorn, VIC, 3122, Australia} 
\affil{Australian Research Council Centre of Excellence for Gravitational Wave Discovery (OzGrav), Australia} 
\affil{Australian Research Council Centre of Excellence for All-Sky Astrophysics in 3 Dimensions (ASTRO-3D), Australia}

\author[0000-0001-7457-8487]{Kentaro Nagamine}
\affiliation{Department of Earth and Space Science, Graduate School of Science, Osaka University, Toyonaka, Osaka, 560-0043, Japan}
\affiliation{Kavli IPMU (WPI), UTIAS, The University of Tokyo, Kashiwa, Chiba 277-8583, Japan}
\affiliation{Department of Physics \& Astronomy, University of Nevada Las Vegas, Las Vegas, NV 89154, USA}

\author[0000-0001-5310-4186]{Jielai Zhang}
\affil{Centre for Astrophysics and Supercomputing, Swinburne University of Technology, Hawthorn, VIC, 3122, Australia} 
\affil{Australian Research Council Centre of Excellence for Gravitational Wave Discovery (OzGrav), Australia}



\begin{abstract}
The repeating fast radio burst FRB20190520B is an anomaly of the FRB population thanks to its high dispersion measure (DM$=1205\,\dmunits$) despite its low redshift of $z_\mathrm{frb}=0.241$. This excess has been attributed to a large host contribution of $\dmhost\approx 900\,\dmunits$, far larger than any other known FRB. In this paper, we describe spectroscopic observations of the FRB20190520B field obtained as part of the FLIMFLAM survey, which yielded 701 galaxy redshifts in the field. We find multiple foreground galaxy groups and clusters, for which we then estimated halo masses by comparing their richness with numerical simulations. We discover two separate $\mhalo >10^{14}\,\msun$ galaxy clusters, at $z=0.1867$ and $z=0.2170$, respectively, that are directly intersected by the FRB sightline within their characteristic {halo} radius \rvir{}. Subtracting off their estimated DM contributions as well that of the diffuse intergalactic medium, we estimate a host contribution of $\dmhost=430^{+140}_{-220}\,\dmunits$ or $\dmhost=280^{+140}_{-170}\,\dmunits$ (observed frame) depending on whether we assume the halo gas extends to $\rvir$ or $2\times\rvir$. This significantly smaller \dmhost{} --- no longer the largest known value --- is now consistent with H$\alpha$ emission measures of the host galaxy without invoking unusually high gas temperatures. {Combined with the observed FRB scattering timescale, we estimate the turbulent fluctuation and geometric amplification factor of the scattering layer to be $\effgee\approx4.5 - 11\,(\mathrm{pc^2\;km})^{-1/3}$, suggesting most of the gas is close to the FRB host}. This result illustrates the importance of incorporating foreground data for FRB analyses, both for understanding the nature of FRBs and to realize their potential as a cosmological probe.
\end{abstract}



\section{Introduction} \label{sec:intro}

Fast radio bursts (FRBs) are phenomena that have excited tremendous interest, not just because of the enigmatic
nature of their source engines, but also because their frequency sweeps encode information on the integrated 
free electron column density along their lines-of-sight.
This is usually quantified through the dispersion measure, $\mathrm{DM} = \int n_e(l) \,\mathrm{d}l$, 
where $n_e(l)$ is the free electron density along the line-of-sight path $l$.

Among the sample of FRBs that have been localized at the time of writing, \thisfrb{} ranks among the most notable.
First reported by \citet{niu:2022}, it was discovered as a sequence of repeating bursts by the Five-Hundred Aperture Spherical Radio Telescope
\citep[FAST;][]{nan:2011,li:2018} that was subsequently localized with the Karl.\ G.\ Jansky Very Large Array \citep[VLA;][]{law:2018} to the Equatorial J2000 coordinates (RA, Dec)=(16:02:04.261, -11:17:17.35). Follow-up optical imaging and spectroscopy revealed a host galaxy, J160204.31-111718.5 (hereafter referred to as HG190520), that was 
associated with the FRB at high confidence. This
galaxy was measured to have a spectroscopic redshift of $\zfrb=0.241$, and also has an associated persistent radio source.

With a total measured dispersion measure of $\dmfrb = 1204.7\pm 4.0\,\dmunits$ \citep{niu:2022}, \thisfrb{} has a DM well in excess of the 
value expected given its redshift
and the Macquart Relation \citep{macquart:2020}: assuming a Milky Way contribution of $\dmmw\sim 100\,\dmunits$ and a mean 
IGM contribution of $\dmaigm \sim 300\,\dmunits$ (using the rough approximations of \citealt{ioka:2003} and \citealt{inoue:2004}), \thisfrb{} exhibits a DM over twice that expected given its redshift.
This was attributed to a large host contribution of $\dmhost = 903^{+72}_{-111}\,\dmunits$ (observed frame; \citealt{niu:2022}), which 
makes it by far the largest host DM value of any known FRB prior to
this current analysis.

While \citet{niu:2022} and \citet{ocker:2022} concluded that no foreground galaxies 
were likely to contribute to the foreground DM, they based this conclusion on a single pointing of Keck/LRIS observations
which was limited to $\sim 2-3\arcmin$ from the FRB sightline.
This would be adequate to reveal, for example, the influence of a foreground galaxy 
at $z=0.15$ with a halo mass $\mhalo\sim 10^{12}\,\msun$ since its characteristic
radius of $\rvir\approx 240\,\kpc$ would extend 1.5$\arcmin$. 
However, a modestly more massive foreground halo with, say, $\mhalo\sim 3\times 10^{13}\,\msun$,
at the same redshift would have a $\rvir$ extending to $4.7\arcmin$ from the
halo center. This would be outside the field of the original optical observations, and would require wide field-of-view multiplexed spectroscopy to characterize the multiple member galaxies. 
Wider observations than originally achieved in the discovery papers are therefore
needed to rule out significant foreground contributions to the large DM of \thisfrb{}.

{In this paper, we describe wide-field multiplexed spectroscopic data we have obtained in the 
\thisfrb{} field as part of the FRB Line-of-sight Ionization Measurement From Lightcone AAOmega Mapping (FLIMFLAM) Survey \citep{lee:2022}. 
This survey, which is carried out on the 2dF/AAOmega fiber spectrograph on the 3.9m Anglo-Australian Telescope (AAT)
in Siding Spring, Australia, is designed to observed large numbers of galaxy redshifts in the foregrounds of 
localized FRBs in order to map the large-scale structure in the foreground. 
As shown by \citet{simha:2020}, this would allow us to separate the various components that make up the DM observed in FRBs. 
We then describe the group-finding approach applied to the spectroscopic data using a commonly-used friends-of-friends (FoF) algorithm to identify galaxy groups and clusters within the \thisfrb{} field.  
The group/cluster halo masses are then estimated by comparing with forward models of group/cluster richness derived from cosmological $N$-body simulations.
Next, we model the implied DM from both the 
foreground halos and diffuse IGM, yielding updated estimates for the \thisfrb{} DM host contribution. 
Finally, we discuss the implications of the new
host estimate in context of the observed host galaxy H$\alpha$ emission and FRB scattering.}
In a separate paper, \citet{simha:2023} had analyzed 4 other sightlines also shown to exhibit an 
excess \dmfrb{} given their redshift, but \thisfrb{} was deemed such an unusual object that it merited a separate analysis and paper.

In this paper, we use the term `groups' to refer to both groups and clusters when we are agnostic as to their underlying halo masses, 
but use `cluster' to refer specifically to objects with $M_\mathrm{halo} \geq 10^{14}\,M_\odot$. We use a concordance $\Lambda$CDM cosmology with $\sigma_8 = 0.829$, $H_0 = 67.3\,\mathrm{km}\, \mathrm{s}^{-1} \mathrm{Mpc}^{-1}$,
$h=H_0/(100\,\mathrm{km}\, \mathrm{s}^{-1} \mathrm{Mpc}^{-1})=0.673$, $ \Omega_\Lambda = 0.685$, $\Omega_m = 0.315$,  
$\Omega_b = 0.0487$, and $n = 0.96$.

\section{Observations} \label{sec:obs}

% Viz_FRB190520_Grps.ipynb
\begin{figure}
\includegraphics[width=0.5\textwidth]{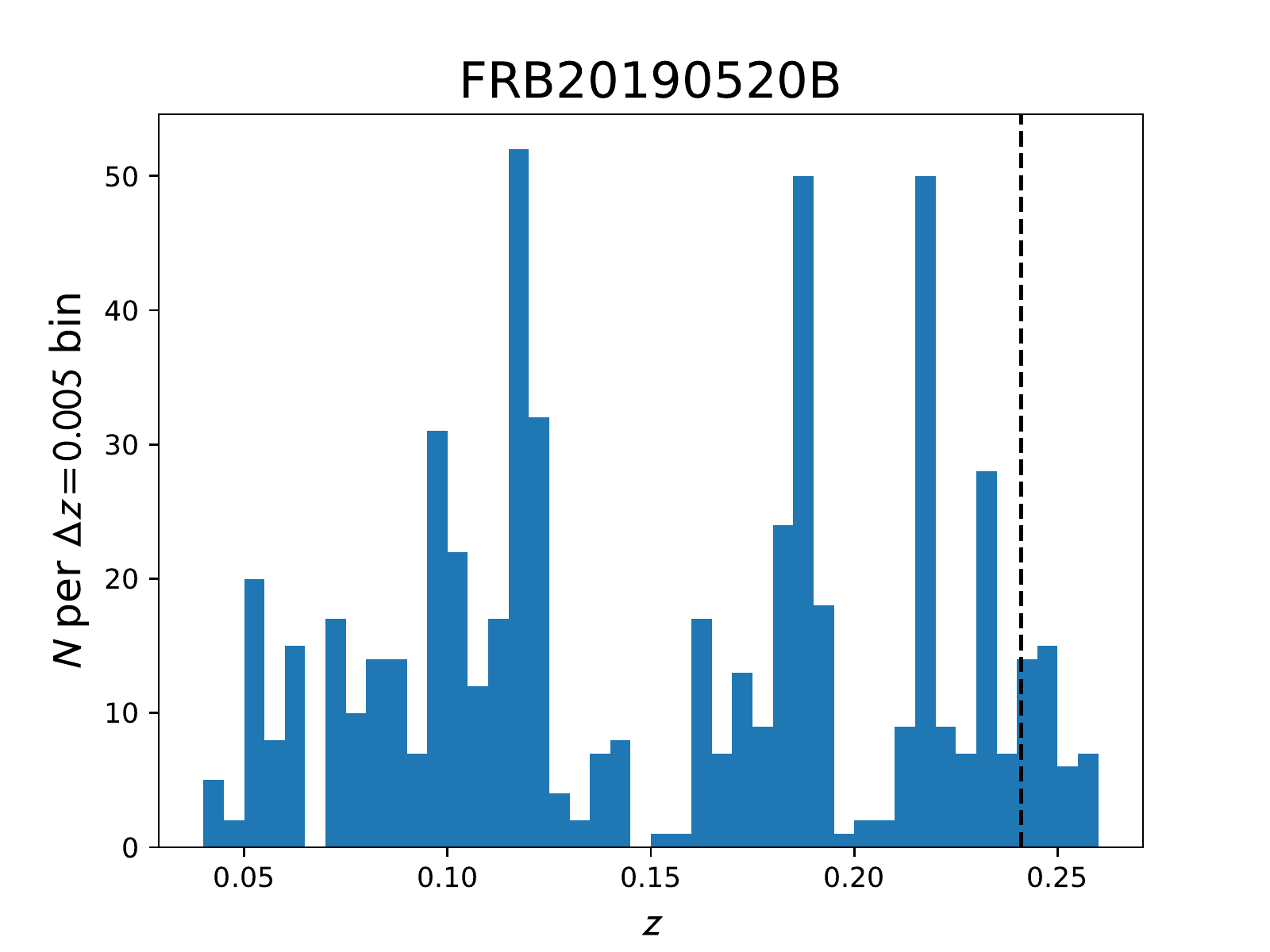}
\caption{\label{fig:zdist}
The galaxy redshift distribution from our AAT/2dF-AAOmega spectroscopic observations, which targeted an approximately 2 degree field of view centered on \thisfrb{}.
The FRB redshift ($\zfrb=0.241$) is indicated by the vertical-dashed line.
}
\end{figure}

In order to characterize the foreground contributions to the \thisfrb{} dispersion measure (DM), we carried out a sequence of observations as part of the broader FLIMFLAM survey \citep{lee:2022}. 
Since \thisfrb{} was immediately reported to have an anomalously high \dmhost{} by \citet{niu:2022}, 
it was deemed to be of sufficient scientific interest to constrain the foreground contributions
of \thisfrb{}, and
therefore it was added to the FLIMFLAM observation list as a special target.

\thisfrb{} was not, to our knowledge, within the footprint of publicly-available imaging surveys, therefore on 2021 August 8 we imaged the field with a single pointing of the Dark Energy Camera (DECam) in $r$-band.
The data was then reduced using the standard pipeline provided by the observatory, and a galaxy catalog generated using \texttt{SourceExtractor} \citep{bertin:1996} after excluding point sources as stars.
Since the DECam imaging preceded the spectroscopic run with AAT/AAOmega by only one month, at the time of spectroscopic target selection we had only preliminary image reductions that did not have a
well-characterized magnitude zero-point.
For the spectroscopic target selection, we therefore defined an arbitrary magnitude cut in the $r$-band to select approximately 1500 targets within the 2 degree diameter (i.e.\ 3.14 deg$^2$) footprint of AAT/AAOmega. 
This was intended to target an areal density of galaxies that should, on average, correspond approximately to a magnitude limit of $r \approx 19.4$\,mag, which is our magnitude limit for FRBs at $z\approx 0.2$ (c.f.\ 
\citealt{simha:2023}).
However, this selection was later found to actually correspond to an apparent depth of $r=19.1$\,mag or unextincted magnitude of $r_\dered = 18.4$\,mag based on
corrections using the \citet{schlafly:2011} dust map\footnote{For notational clarity, all magnitudes quoted subsequently in this paper will implicitly be corrected for dust extinction magnitudes unless stated otherwise.}.
This field is thus overdense in galaxies given the relative shallowness of the magnitude threshold, as we shall see later. We note that there is significant variation ($\sim 0.4$ mag) in the extinction across our
3.1 deg$^2$ field which makes it challenging to compare the number counts with simulations, but our forward model described later should be accurate in the vicinity of the FRB position. 

Using \texttt{configure} \citep{miszalski:2006}, the standard plate configuration software provided by the observatory,
we designed 5 plates filled with approximately 300 galaxies each across the 3.1 sq deg 2dF footprint, of which 3 were observed with AAOmega on UT 2021 Sept 7-9. 
The observing setup was the 580V grating blazed at 485nm on the blue camera, while the red camera used the 385R grating blazed at 720nm. 
In combination with the 570nm dichroic, this allowed a continuous spectral coverage across 380nm-880nm with a spectral resolving power of $R\approx 1300$.   
Each plate received 4500s-5400s of on-sky exposure; the first plate was observed in sub-optimal seeing of $3-3.5\arcsec$ on 2021 Sept 7, but the subsequent two plates were observed under nominal conditions in the following nights with $1.6-1.8\arcsec$ seeing.

The raw data was reduced using a version of the \verb|2dFDR| data reduction package kindly provided by the OzDES collaboration
\citep{yuan:2015,childress:2017}. 
We then ran the \verb|MARZ| software \citep{hinton:2016} on the spectra to measure spectroscopic
redshifts, which were then visually confirmed. 
Redshift identifications that appeared to be a reasonable match to the spectral templates were assigned a 
quality operator (QOP) flag of 3, while high-confidence redshifts with multiple high signal-to-noise features 
were assigned QOP$=4$.
There were 701 galaxies that had QOP$=(3 \lor 4)$, which we consider to be `successful' redshifts and will treat
identically in the subsequent analysis. 
The galaxy redshift histogram is shown in Figure~\ref{fig:zdist}, which shows distinct redshift peaks that are suggestive of overdensities
within the field. This motivates us to apply group-finding algorithms to search for galaxy groups or clusters within the sample.

\begin{figure*}\centering
\begin{interactive}{js}{frb190520_field.html}
\includegraphics[width=0.9\textwidth]{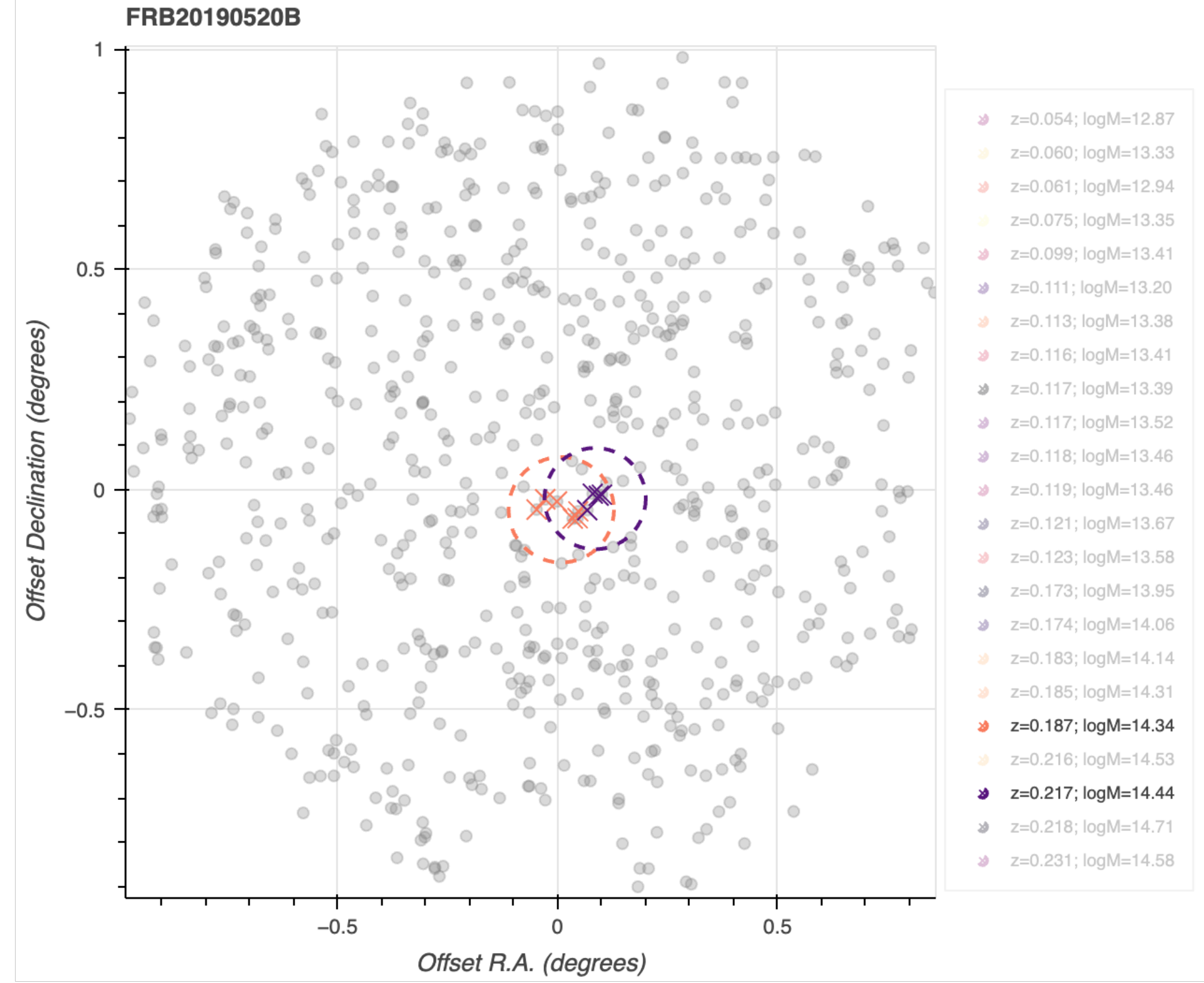}
\end{interactive}
\caption{\label{fig:field_bokeh}
Preview of an interactive visualization of the FLIMFLAM foreground redshifts in the \thisfrb{} field (\href{https://member.ipmu.jp/kg.lee/frb190520_field.html}{click on this URL for full version}). The gray points indicate galaxies with measured
spectroscopic redshifts, with their positions on the sky shown as a relative angular offset from the \thisfrb{} coordinate (i.e.\ the FRB is at [0,0]). In the online version, 
pan-dragging and zooming is supported (tool bar on top), while
hovering over an individual galaxy with the mouse will show the coordinate and redshift.
Clicking on the legend on the right will highlight galaxy groups labeled by redshift, with group members marked with '$\times$' 
and the physical extent of \rvir{} indicated by a dashed circle. In the static preview, we have highlighted the clusters at $\zgrp=0.1867$ and $\zgrp=0.2170$
that are believed to provide large contributions to \dmhalos. Full version will be available in the online published journal.
}
\end{figure*}

% These numbers are in GroupPropertiesFRB190520.ipynb
During visual inspection of the spectra, we found 122 confirmed stars compared to 701 galaxies, or a stellar contamination rate of 17.4\%. 
This is relatively high compared to the other fields observed by FLIMFLAM, presumably due to the low Galactic latitude 
($b= 29$\,deg)
and imperfect star-galaxy separation. 
If we assume that the stellar contamination rate is same between the spectroscopically confirmed targets and those that could not be assigned a
conclusive redshift, then the spectroscopic completeness success rate of our galaxy sample is 701 out of 1260 possible galaxy targets, 
or 56.6\%.

\section{Halos} \label{sec:groupfinding}
We applied an anisotropic
friends-of-friends (FoF) group-finding algorithm on our \thisfrb{} spectroscopic data, kindly provided by Elmo Tempel \citep{tago:2008,tempel:2012,tempel:2014a}. 
While standard FoF algorithms, such as those used to identify dark matter halos in $N$-body simulations, typically adopt the same linking length in all 3 dimensions, Tempel's code allows for a longer linking length in the radial dimension 
in order to account for redshift-space distortions, i.e. the ``Fingers of God''. 

% fof_pars_frb190520.ipynb
This finder assumes a transverse linking length, $\dll$, which varies as a function of redshift, $z$, in the following way:

\begin{equation}
\dll(z) = d_\mathrm{LL,0}[1 + a \,\mathrm{arctan}(z/z_*)],
\end{equation}

\noindent where $d_\mathrm{LL,0}$ is the linking length at a fiducial redshift, while
$a$ and $z_*$ govern the redshift evolution.
This redshift dependence accounts for the artificial decreasing number of galaxies as a function of redshift, within a flux-limited survey. 
We then set the radial linking length, $\dpar(z)$, to be proportional to $\dll(z)$. 
The final parameters used for group-finding in this paper are: 
$\dll=0.35\,\hMpc$, $a=0.75$, $z_*=0.1$, and $\dpar/\dll=10$. 
Note that we adopt a linking length which is larger than used in \citet{simha:2023} 
in order to account for our sparser sample of redshifts, which is nearly a factor of two
lower space density than in \citet{simha:2023}. 

The resulting group catalog has groups with richness or multiplicity (i.e.\ number of identified galaxies) as low as $\nrich=2$, but we limit ourselves to $\nrich=4$ to have a more robust sample\footnote{We use a slightly more aggressive richness cut than in \citet{simha:2023} because we will not use the virial theorem to estimate mass.}. 
These are listed in Table~\ref{tab:groups}, although note that we omit groups with $\zgrp > \zfrb$. We do not find any group with $\zgrp \approx \zfrb$, which can be plausibly associated with the FRB host. 

In Figure ~\ref{fig:field_bokeh}, we present an interactive plot\footnote{Made using Bokeh: \url{http://www.bokeh.pydata.org}.} that shows the position of the galaxies as well as identified groups within the FRB field.

\begin{deluxetable*}{lccccccccc}
\tablecaption{\label{tab:groups}
Detected Galaxy Groups and Clusters in the \thisfrb{} Field. 
}
\tablehead{ \multirow{2}{*}{ID} & \multirow{2}{*}{$\zgrp$} & \colhead{RA\tablenotemark{a}} & \colhead{Dec\tablenotemark{a}} & 
\colhead{$b_\perp$\tablenotemark{b}} & \colhead{$b_\perp$\tablenotemark{b}} & \multirow{2}{*}{\nrich}\tablenotemark{c} & \multirow{2}{*}{$\log_{10}\!\left(\frac{\mhalo}{\msun}\right)$} & \colhead{\rvir{}\tablenotemark{d}} & \colhead{$b_\perp$\tablenotemark{b}} \\[-8pt]
  &  & \colhead{(deg)} & \colhead{(deg)} & \colhead{(arcmin)} & \colhead{(kpc)} &  &
   & \colhead{(kpc)} & \colhead{(\rvir)}
}
\startdata
\vspace*{3pt}
2403390701118753 & 0.0536 & 240.3391 & -11.1875 & 12.13 & 789.1 & 4 & $ 12.87^{+0.21}_{-0.90}$  & 453  & 1.74  \\[3pt]
2400721001053090 & 0.0605 & 240.0721 & -10.5309 & 52.48 & 3820.3 & 7 & $ 13.33^{+0.10}_{-0.10}$ & 644  & 5.93  \\[3pt]
2407627701151079 & 0.0608 & 240.7628 & -11.5108 & 19.65 & 1437.7 & 4 & $ 12.94^{+0.28}_{-0.23}$ & 478  & 3.01  \\[3pt]
2402176201159096 & 0.0745 & 240.2176 & -11.5910 & 25.33 & 2236.3 & 6 & $ 13.35^{+0.15}_{-0.28}$ & 654  & 3.42  \\[3pt]
2399925001187499 & 0.0991 & 239.9925 & -11.8750 & 46.83 & 5342.0 & 5 & $ 13.41^{+0.16}_{-0.46}$ & 685  & 7.80  \\[3pt]
2404259001164001 & 0.1114 & 240.4259 & -11.6400 & 21.79 & 2753.9 & 4 & $ 13.20^{+0.34}_{-0.37}$ & 583  & 4.72  \\[3pt]
2404782701043103 & 0.1133 & 240.4783 & -10.4310 & 51.48 & 6602.6 & 4 & $ 13.38^{+0.23}_{-0.58}$ & 670  & 9.86  \\[3pt]
2401932701109639 & 0.1163 & 240.1933 & -11.0964 & 22.30 & 2925.7 & 4 & $ 13.41^{+0.25}_{-0.59}$ & 685  & 4.27  \\[3pt]
2395862301164344 & 0.1166 & 239.5862 & -11.6434 & 58.78 & 7728.4 & 4 & $ 13.39^{+0.25}_{-0.56}$ & 675  & 11.45 \\[3pt]
2404329301079377 & 0.1169 & 240.4329 & -10.7938 & 30.08 & 3965.7 & 5 & $ 13.52^{+0.19}_{-0.43}$ & 746  & 5.32  \\[3pt]
2405894601098030 & 0.1176 & 240.5895 & -10.9803 & 18.95 & 2509.3 & 4 & $ 13.46^{+0.21}_{-0.63}$ & 712  & 3.52  \\[3pt]
2406789301111492 & 0.1189 & 240.6789 & -11.1149 & 14.07 & 1881.9 & 4 & $ 13.46^{+0.21}_{-0.58}$ & 712  & 2.64  \\[3pt]
2409852201173268 & 0.1211 & 240.9852 & -11.7327 & 38.30 & 5203.4 & 6 & $ 13.67^{+0.17}_{-0.30}$ & 836  & 6.22  \\[3pt]
2408283601132797 & 0.1228 & 240.8284 & -11.3280 & 18.43 & 2533.4 & 5 & $ 13.58^{+0.20}_{-0.48}$ & 781  & 3.25  \\[3pt]
2408321001051411 & 0.1727 & 240.8321 & -10.5141 & 50.00 & 9140.7 & 4 & $ 13.95^{+0.23}_{-0.47}$ & 1037 & 8.81  \\[3pt]
2411417701121786 & 0.1741 & 241.1418 & -11.2179 & 36.96 & 6803.0 & 5 & $ 14.06^{+0.20}_{-0.30}$ & 1128 & 6.03  \\[3pt]
2402401701215090 & 0.1833 & 240.2402 & -12.1509 & 54.27 & 10409.6 & 4 & $ 14.14^{+0.19}_{-0.83}$& 1200 & 8.68  \\[3pt]
2405098001170259 & 0.1851 & 240.5098 & -11.7026 & 24.87 & 4807.9 & 6 & $ 14.31^{+0.19}_{-0.30}$ & 1367 & 3.52  \\[3pt]
2405265701133378 & 0.1867 & 240.5266 & -11.3338 & 2.79 & 542.4 & 6 & $ 14.34^{+0.18}_{-0.28}$   & 1399 & 0.39  \\[3pt]
2401408901101955 & 0.2164 & 240.1409 & -11.0195 & 27.42 & 5989.1 & 5 & $ 14.53^{+0.19}_{-0.31}$ & 1619 & 3.70  \\[3pt]
2406051301130845 & 0.2170 & 240.6051 & -11.3084 & 5.28 & 1155.5 & 4 & $ 14.44^{+0.20}_{-0.40}$  & 1511 & 0.76  \\[3pt]
2407679601092642 & 0.2175 & 240.7680 & -10.9264 & 26.23 & 5750.6 & 7 & $ 14.71^{+0.18}_{-0.25}$ & 1858 & 3.09  \\[3pt]
2405885701164888 & 0.2315 & 240.5886 & -11.6489 & 22.04 & 5065.3 & 4 & $ 14.58^{+0.22}_{-0.43}$ & 1682 & 3.01 
\enddata
\tablenotetext{a}{R.A.\ and declination in Equatorial J2000 coordinates.} 
\tablenotetext{b}{Impact parameter from the FRB sightline, in various units.}
\tablenotetext{c}{Observed group richness.}
\tablenotetext{d}{Halo characteristic radius at which matter density is 200$\times$ critical density.}
\end{deluxetable*}

\subsection{Estimating Group/Cluster Masses} \label{sec:groupmasses}
In \citet{simha:2023}, we adopted dynamical halo masses estimated using the virial theorem applied to the projected size and velocity dispersion of the galaxy groups. 
For the foreground sample of \thisfrb, however, the observations were much shallower and the spectroscopic success rate was considerably lower than equivalent datasets
in \citet{simha:2023}. 
For example, \citet{simha:2023} had 1493 galaxy redshifts for the FRB20190714A field which is at $\zfrb=0.2365$, whereas for \thisfrb{} at the similar redshift of $\zfrb=0.241$ we only have 701 successful galaxy redshifts.
We therefore consider the \thisfrb{} sample too sparse for reliable application of the virial theorem for group/cluster mass estimation. 

Instead, we use the group richness, \nrich, combined with a forward-modeling approach based on semi-analytic models of galaxy formation.
Specifically, we use the \citet{henriques:2015} lightcone catalogs that were generated by
applying the ``Munich'' semi-analytic galaxy formation model \citep{guo:2011} to the $L=500\,\hMpc$ Millennium $N$-body simulation 
\citep{springel:2005}. 
In particular, we use the ``all-sky'' catalogs that are designed to simulate footprints covering $4\pi$ sr in order to maximize the number of simulated groups and clusters. 
We queried the simulation SQL database\footnote{\url{http://gavo.mpa-garching.mpg.de/MyMillennium}} to download $z<0.25$ cluster catalogs with $\lgmh \geq 14.2$ from over the full sky, whereas for the halo mass range $13.0 \leq \lgmh < 14.2$ 
a subsample over a footprint of 400 deg$^2$ was deemed to provide a sufficient sample size to probe the diversity of the groups without having
to download an excessively large file.
This simulated group catalog includes effects such as $K$-corrected photometry based on the realistic galaxy spectral energy distributions 
(including the effect of dust), redshift space distortions, and the distance modulus (i.e. increasingly faint magnitudes as a function of redshift).

\begin{figure}
% BuildClusterMassEstimator.ipynb
\includegraphics[width=0.49\textwidth]{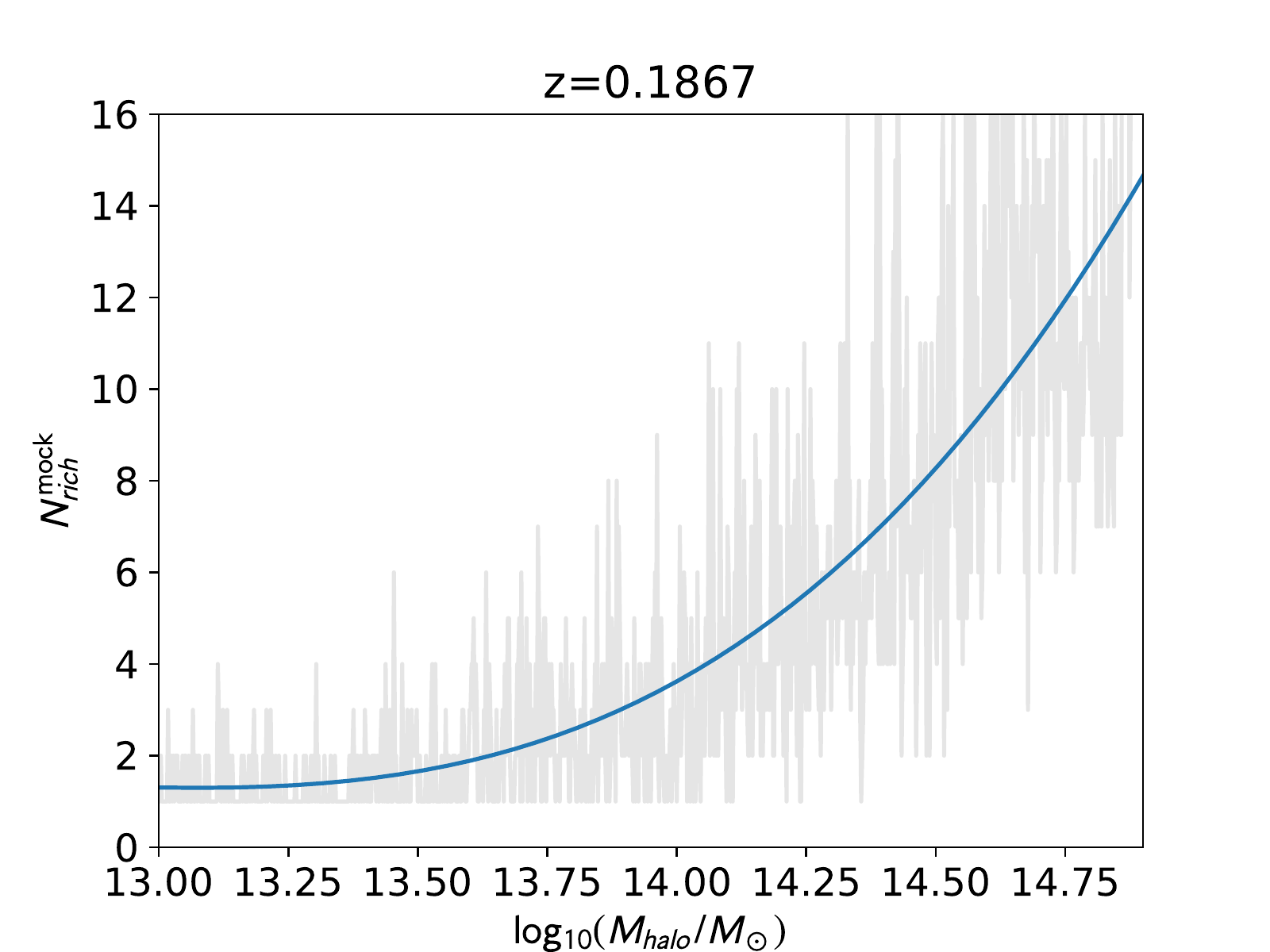}
\caption{\label{fig:nrich}
An example of our halo mass estimation model using the group richness, built from mock group/clusters from the Millennium simulation.
The light grey curve show the richness of an ensemble of mock groups selected at the redshift of our detected group at $\zgrp=0.1867$,
after incorporating the incompleteness and magnitude threshold of our data. 
The curve is the spline fit to this ensemble of mock groups, which is then used to estimate the halo masses
as a function of the group richness.
}
\end{figure}

Using these mock catalogs, we build forward models matched to the observed redshift, \zgrp, of each detected group in our sample. 
First, we selected groups/clusters from the mock catalog within $\zgrp \pm 0.005$, and applied the same magnitude cut as that corresponding to our field center, $r\leq 18.42$ (with 
a small distance modulus correction to account for the finite width of the $\Delta z=0.005$ redshift selection bin). 
This step yields \nrichtrue{} group members that would be observed within our observing setup if we had achieved 100\% spectroscopic completeness. 
We then downsample each group to our actual observed completeness by drawing a random Poisson variate with a mean of $\mu = \fobs\,\nrichtrue$, 
where $\fobs=56.6\%$ is the completeness of our \thisfrb{} spectroscopy.
The selected group galaxies are used to compute the observed mock richness, $\nrichmock = \mu\,\nrichtrue$, for that simulated group. 
In Figure~\ref{fig:nrich}, the gray lines show \nrichmock{} computed as a function of $\lgmhnu$ for an ensemble of simulated groups
selected to be at the same redshift as one of our observed groups. 
The scatter originates from both the intrinsic diversity of group properties at fixed halo mass, and also from the
stochasticity induced by the Poisson sampling at our low spectroscopic completeness. 
We then fit a spline function to obtain \nrichmock{} as a function of $\lgmhnu$.
While there are gaps in the sample of mock groups at certain masses, especially at the massive end, the 
spline fit appears to be a reasonable model for $\nrichmock$ as a function of $\lgmhnu$.

To estimate the halo masses, we then use the standard Chi-squared minimization methodology:

\begin{equation}\label{eq:chisq}
\chi^2 =\frac{(\nrich - \nrichmock)^2 }{\sigma_N^2} ,
\end{equation}

\noindent where $\sigma_N = \sqrt{\nrich}$ is the observational errors on $\nrich$, which we estimate to be simply the Poisson uncertainty. 
We evaluate this on a grid of halo masses in the range $13.0 < \lgmh < 14.9$ using the spline-fitted model for \nrichmock{} described above obtained for each redshift.
The best-fit halo mass is given by the minimum $\chi^2$, while we also estimate the 68th percentile uncertainties by evaluating the halo masses
at $\Delta \chi^2 = 1$.
The estimated halo masses are listed in Table~\ref{tab:groups}, along with the corresponding \rvir{}, the characteristic
halo radius within which the halo matter density is $200\times$ the mean density of the Universe at that redshift.

Notably, we find several foreground galaxy clusters with halo masses of $\lgmh>14$. Two of these clusters
 are intersected by the FRB sightline within their characteristic cluster radius:
(1) at $\zgrp = 0.1867$, a galaxy cluster with $\lgmh = 14.34^{+0.18}_{-0.28}$ lies at a mere 2.79$\arcmin$
or $b_\perp = 542\,\kpc$ 
from the FRB sightline, which corresponds to 0.388$\times \rvir$ and;
(2) 
another, separate, cluster is at $z=0.2170$ with an estimated halo mass of $\lgmh = 14.44^{+0.20}_{-0.40}$,
intersecting the sightline at 0.765$\times \rvir$.

\begin{figure} % group_analysis/Tweak_GroupSelection.ipynb
\includegraphics[width=0.5\textwidth]{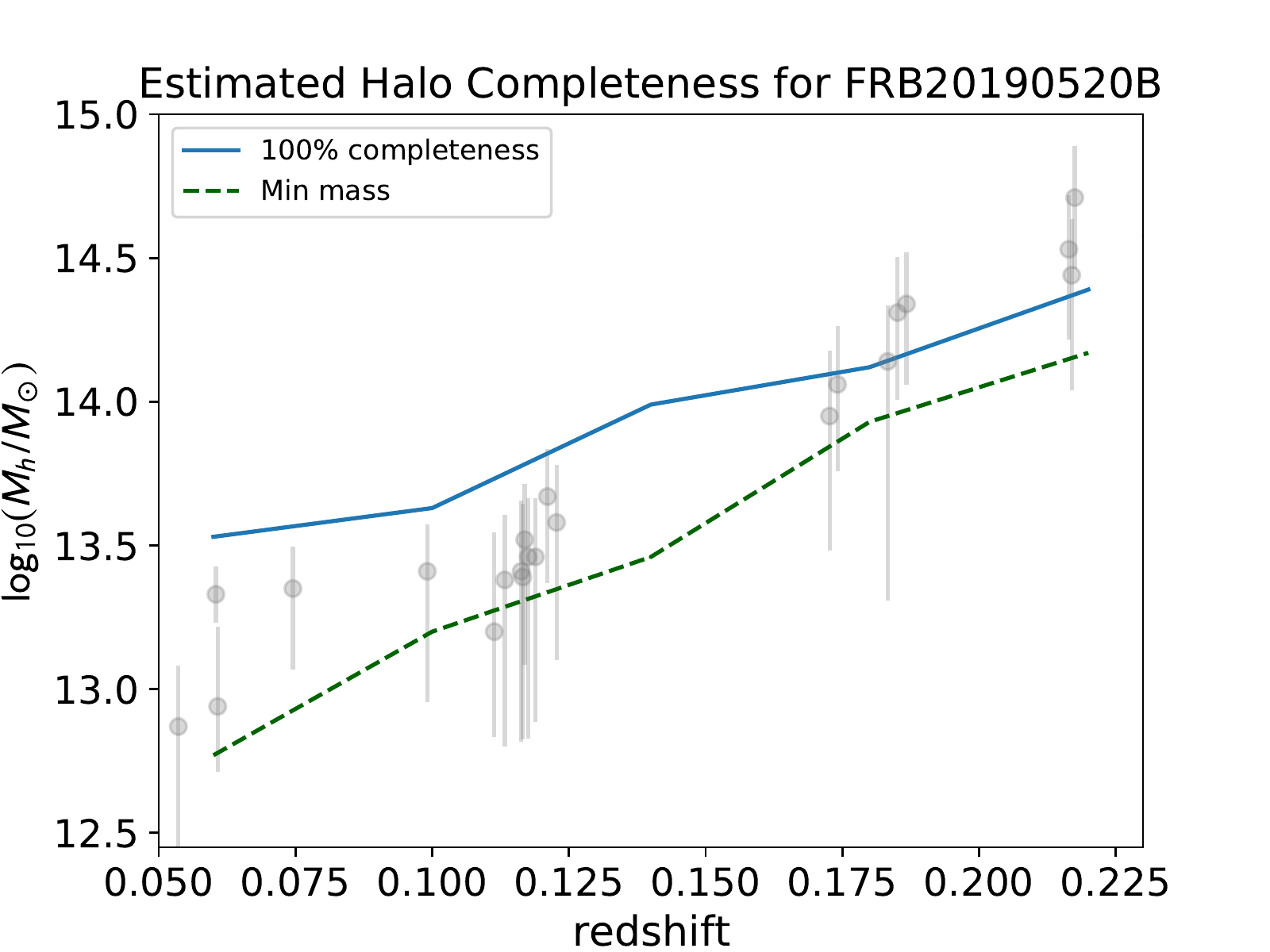}
\caption{\label{fig:mass_complete}
Data points show the redshift and estimated halo masses of the detected groups from our \thisfrb{} spectroscopic data. 
Curves show the estimated mass completeness of simulated groups and clusters as a function of redshift, assuming the observational depth and completeness
of our data.
The dashed curve indicates the minimum mass at which groups are detectable with $\nrich \geq 4$ at the given redshift, while
the solid curve indicates the mass at which we expect to be 100\% complete.  
}
\end{figure}

At first glance, it might appear that there are preferentially more low-mass groups at the low-redshift
end of our spectroscopic data, while the more massive clusters lie preferentially toward the higher redshifts. 
However, this is a selection effect in that only more massive groups or clusters are detectable as multiple galaxy members with our relatively
shallow magnitude threshold and low completeness.
This is shown in Figure~\ref{fig:mass_complete}, in which we use the lightcone galaxy catalogs of \citet{henriques:2015} to derive the our expected completeness
of $\nrich \geq 4$ groups and clusters as a function of halo mass and redshift. This takes into account the estimated observational depth and incompleteness, as well 
as the intrinsic variance in the number of member galaxies for each halo mass. 
The masses that can be selected do indeed increase as a function of redshift, consistent with what we see in our data.
While we are confident that we have ruled out massive clusters at $z<0.15$ in our field, 
we suspect that more complete spectroscopic observations would reveal multiple lower-mass groups associated with the clusters detected at $z>0.17$.
Indeed, our detected clusters are not isolated, with multiple entities detected at similar redshifts --- indicative of true overdensities
and related structures in the cosmic web.

\section{Analysis}

In this paper, we adopt the observed DM value of $\dmfrb=(1205 \pm 4)\, \dmunits$ for \thisfrb{}, as reported by \citet{niu:2022}.
We then decompose the total observed DM contribution for \thisfrb{} into several components:

\begin{equation}\label{eq:dm_all}
\dmfrb = \dmmw + \dmhalos + \dmigm + \dmhost,
\end{equation}

\noindent where \dmmw{} is the contribution from the Milky Way, \dmhalos{} is the summed contribution from individual halos\footnote{We use \dmhalo{} to refer to 
individual halo contributions, while \dmhalos{} (note plural in subscript) is the aggregate quantity from all halos.} that intersect the sightline, 
\dmigm{} is the contribution from the diffuse intergalactic medium (IGM) outside of halos, and \dmhost{} is the combined contribution from the host galaxy and FRB engine. 
Note that the notation \dmigm{} is sometimes used to refer to the sum of
\dmhalos{} and \dmigm{} in the literature, but in this paper we explicitly separate out the IGM and halo contributions.
For the Milky Way component, we use the same
estimate by \citet{niu:2022} of $\dmmw= 113 \,\dmunits$ for the disk and halo contribution from the Milky Way, respectively. The disk contribution was {estimated by averaging the NE2001 \citep{cordes:2002} and YMW16 \citep{yao:2017} disk models.
The Galactic halo contribution, on the other hand, was estimated using models from
\citet{prochaska:2019}, \citet{ravi:2023}, and \citet{cook:2023}.}

In the following subsections, we will assess the contributions of \dmhalos{} and \dmigm{} based on our 
observational data.

\subsection{Foreground Halo Contributions}\label{sec:halos}
We have now established that the \thisfrb{} sightline directly intersects within $<1\,\rvir$ of
at least two galaxy clusters in the foreground. 
In order to calculate the \dmhalos{} contribution, we need to make assumptions about the 
 {distribution of free electrons in the circum-halo medium of these halos.}

% GenLatexTable_DMhalos_FRB190520
\begin{deluxetable*}{lccccccccc}[ht!] 
% There is a dummy column between b and DM_halo to allow a break in the \cline
\tablecaption{\label{tab:dmhalo}
Foreground \dmhalo{} Contributors}
\tablehead{\multirow{2}{*}{ID} & \multirow{2}{*}{$z$} & \colhead{RA\tablenotemark{a}}   & \colhead{Dec\tablenotemark{a}} & \multirow{2}{*}{\lgmh} & \multicolumn{2}{c}{$b_\perp$} & & \multicolumn{2}{c}{\dmhalo{} ($\dmunits$)\tablenotemark{b} } \\[-9pt] \noalign{\vskip 8pt} \cline{6-7} \cline{9-10}  &   & \colhead{(deg)} & \colhead{(deg)}  &    &   (kpc)   & (\rvir) & & \modelone & \modeltwo
}
\startdata
2403390701118753 & 0.0536 & 240.3391 & -11.1875 & $12.87^{+0.21}_{-0.90}$ & 789 & 1.58 & & $0^{+20}_{-0}$ & $20^{+65}_{-12}$ \\[3pt]
2405167901131499 & 0.1862 & 240.5168 & -11.3150 & $11.78^{+0.30}_{-0.30}$ & 313 & 1.57 & & $0^{+0}_{-0}$ & $10^{+17}_{-10}$ \\[3pt]
2405265701133378 & 0.1867 & 240.5266 & -11.3338 & $14.34^{+0.18}_{-0.28}$ & 542 & 0.38 & & $300^{+130}_{-81}$ & $350^{+140}_{-87}$ \\[3pt]
2406051301130845 & 0.2170 & 240.6051 & -11.3085 & $14.44^{+0.20}_{-0.40}$ & 1156 & 0.76 & & $110^{+150}_{-50}$ & $180^{+160}_{-52}$ \\[3pt]
\enddata
\tablenotetext{a}{R.A.\ and declination in Equatorial J2000 coordinates} 
\tablenotetext{b}{Contribution to the \thisfrb{} dispersion measure, assuming two different CGM models with \modelone{} and \modeltwo{}. We retain only two significant figures of these results.}
\end{deluxetable*}

{With the assumption that the free electrons trace the fully ionized 
circum-galactic halo gas}, 
we use the same halo gas density profile previously adopted in \citet{simha:2020, simha:2021, simha:2023}, in which the radial profile of the halo baryonic gas density is:

\begin{equation} \label{eq:mnfw}
    \rho_b(y) = \fhot \left(\frac{\Omega_b}{\Omega_m} \right) \left[\frac{\rho_0(\mhalo)}{y^{1-\alpha}\,(y_0 + y)^{2+\alpha}}\right].
\end{equation}

\noindent The expression in square parentheses is a modified Navarro-Frenk-White (\citealt{navarro:1997}; hereafter mNFW) radial halo profile for the matter density, as modified by \citet{prochaska:2019} in order to approximate the properties
of a multi-phase circum-galactic medium (CGM) \citep{maller:2004}. With the central density of $\rho_0$ set by the halo mass \mhalo, it is a function of $y$, the scaled radius parameter (see \citealt{mathews:2017}), while we adopt $\alpha=2$ and $y_0 = 2$ in this analysis. 
The ratio $\Omega_b/\Omega_m=0.157$ is the baryon fraction relative to the overall matter density, 
while \fhot{} determines the amount of the baryons that are present in the hot gas of the halo. 
The truncation radius of the gaseous halo, \rmax{}, is another free parameter of this model.

{
In our model, the \dmhalo{} contributed by a halo intersected at fixed impact parameter
is therefore a function of $\{\mhalo, \rmax, \fhot\}$. 
The uncertainties in \mhalo{} have been explicitly estimated in Section~\ref{sec:groupmasses}
and will be directly taken into account in the subsequent analysis. 
To incorporate some of the uncertainties in $\rmax$ and $\fhot$, however, 
we adopt two different models for the foreground halo contributions that we believe bracket
their plausible range}:
\begin{enumerate}
    \item We truncate the gas halos at $\rmax=\rvir$. For cluster-sized halos ($\mhalo > 10^{14}\,\msun$), we adopt a halo gas fraction of $\fhot=0.90$. This is driven by X-ray constraints on the baryonic gas fractions in intra-cluster media, that suggest that clusters retain essentially
    all their baryonic content thanks to their deep gravitational potentials \citep[e.g.,][]{gonzalez:2013,chiu:2018}. 
    The value of $\fhot=0.90$ assumes that stars and ISM within member galaxies comprise $f_\star\approx 0.1$ of the cluster baryons, with the rest residing in the intra-cluster medium (ICM) gas.
    For lower-mass halos with $\mhalo < 10^{14}\,\msun$, we use the same value of $\fhot=0.75$ that was adopted by \citet{prochaska:2019} and \citet{simha:2020}, which allows for some of the halo gas to have been expelled from within the characteristic halo radius.
    
    \item The truncation radius of the gas halos is now increased to $\rmax=2\,\rvir$. For cluster halos with $\mhalo > 10^{14}\,\msun$, 
    we again assume $\fhot=0.90$. For the lower-mass halos, however, we introduce a baryonic `cavity' to the central parts of the halos, such that we have $\fhot = 0.3$ at $r<\rvir$ and a `pile-up' of evacuated baryons of $\fhot = 2$ at $\rvir < r < 2\,\rvir$. 
    This is inspired by recent results from hydrodynamical simulations \citep[e.g.,][]{sorini:2022,ayromlou:2022}, that suggest that 
    galaxy or AGN feedback processes can eject baryons from the central regions of galactic halos into the surrounding environment, 
    leaving a reduced baryon fraction in the halo CGM compared with the primordial value.
\end{enumerate}
For brevity, we will refer to the above models as the \modelone{} and \modeltwo{} models, respectively.
{While it is in principle possible for \rmax{} to be larger, the modified NFW declines radially, so the \dmhalo{} contribution converges with increasing \rmax. For example, for a halo intersected at $b_\perp =0.5\,\rvir$, the \dmhalo{} increases by 24\% when \rmax{} is increased from $\rmax=\rvir$ to $\rmax=2\,\rvir$, but the corresponding increase is only about 5\% going from  $\rmax=2\,\rvir$ to $\rmax=3\,\rvir$.}

To compute \dmhalos{}, we first generate a group halo catalog of the FLIMFLAM spectroscopic data, in which 
the groups and clusters detected in Section~\ref{sec:groupfinding} are each treated as individual halos, 
with the masses estimated from Section~\ref{sec:groupmasses}.
We then removed the member galaxies of these groups from the overall spectroscopic catalog, and treat
the remaining field galaxies as individual, lower-mass, halos.

To assign a halo mass to the field galaxies, we first estimated the stellar masses, \mstar{}, by running 
the galaxy population synthesis code \verb|CIGALE| \citep{boquien:2019} on the $griz$ photometry, with
the redshifts fixed by our spectroscopy. 
The stellar masses were then converted into halo masses using the stellar mass-halo mass relationship (SHMR) of 
\citet{moster:2013}.

The halo lists from the field galaxies and the identified groups were collated for the \dmhalo{} calculation, which integrates the line-of-sight segment 
through the gas halo profile
of Equation~\ref{eq:mnfw} assuming the corresponding halo masses, impact parameters from the FRB sightline, and maximum extent of the halo profiles (\rmax). 
We did this as a Monte Carlo calculation in order to take into account the uncertainties in the halo masses. 
For field galaxies, we drew random realizations corresponding to the mean halo mass as well as a standard deviation of 0.3 dex, 
the latter which is a typical halo
mass error taking into account uncertainties in the stellar mass estimation as well as the intrinsic scatter in the SHMR. 
For the groups and clusters, we sample over the halo mass uncertainties listed in Table~\ref{tab:groups}. 
{However, the $\chi^2$ distributions are asymmetric about the minima, therefore we approximate this by sampling from an asymmetric Gaussian distribution
based on the 16th and 84th percentile errors either side of the best-fit value.} 
In other words, we first draw the Gaussian random deviate
with unit standard deviation, and then scale them by the upper or lower error bars depending on whether the draw is positive or negative.
We repeat this exercise $\sim 100$ iterations, and keep track of the resulting individual \dmhalo{} contributions from every iteration. 

In Table~\ref{tab:dmhalo}, we list the foreground halos that provide non-zero contributions to \dmhalos{}. 
The $\zgrp=0.1867$ cluster which is intersected well within its virial radius ($b_\perp \sim 0.4\,\rvir$) provides the largest contribution, 
with $\dmhalo\approx 300\,\dmunits$ and $\dmhalo\approx 350\,\dmunits$ assuming the \modelone{} and \modeltwo{} models, respectively.
There is also a large contribution from the $\zgrp=0.2170$ cluster, which has a slightly larger mass but is intersected through its outskirts ($b_\perp \sim 0.8\,\rvir$).
Because of this, the two different truncation radii provide a relatively greater difference for this cluster, changing from $\dmhalo\approx 110\,\dmunits$ to 
$\dmhalo \approx 180\,\dmunits$ with the increased gas halo radius. The  $\zgrp=0.1867$ cluster, on the other hand, exhibits a smaller fractional change in \dmhalo{} with respect to \rmax{} since the
sightline passes through the central regions of the halo that gives large contributions with less sensitivity to \rmax.

There is also a possible contribution from the lower-redshift galaxy group at $\zgrp=0.0536$, which has a halo mass of $\mhalo \approx 9\times 10^{13}\,\msun$
but is intersected at $b_\perp \approx 1.6\,\rvir$, nominally outside the characteristic halo radius. For the \modelone{} model, this clearly cannot contribute to \dmhalos{}, but does provide a contribution of
$\dmhalo \approx 20\,\dmunits$ for the \modeltwo{} model.
{In this extended model, another small contribution ($\dmhalo \approx 10\,\dmunits$) is provided by an individual galaxy at $z=0.1862$ with a halo mass of $\mhalo \approx 6\times 10^{11}\,\msun$ intersected also at $b_\perp \approx 1.6\,\rvir$.
While it is debatable whether individual group/cluster members should have their
subhalos modeled separately from the main halo, in our case the difference is negligible.}

\subsection{IGM Contribution} \label{sec:igm}
The FLIMFLAM survey was designed to observe numbers of foreground galaxies to act as tracers  for density reconstructions of the large-scale density field, in order to enable precise constraints on the \dmigm{} contribution (e.g., \citealt{simha:2020},
\citealt{lee:2022}).
However, in the case of \thisfrb{} the observations were shallower and more incomplete for its FRB redshift due to the significant dust extinction within the field. 
The \thisfrb{} data was therefore deemed insufficient for density reconstructions using, e.g. methods from \citet{ata:2015}, \citet{kitaura:2021}, or \citet{horowitz:2019}. 
Therefore, instead of a bespoke calculation of \dmigm{} based directly on the observed foreground field, 
in principle we have to settle for
a global estimate of $\dmaigm(\zfrb)$.

However, we do have a catalog of foreground groups and clusters, which we will take into account when trying to estimate \dmaigm.
Again, we use the `all-sky'  \citet{henriques:2015} lightcone catalogs and associated density fields from the Millennium simulation \citep{springel:2005}, 
largely following the methodology described in Section~3.1 of \citet{lee:2022}.
In order to avoid double-counting of the group/cluster contributions in both \dmaigm{} and \dmhalos{}, we `clip' the simulation density field within two grid cells of groups 
and clusters within the lightcone, with the cell values clipped to $\delta \equiv \rho/\langle \rho \rangle -1 =3$. 
Whether or not a group/cluster is clipped from the density field depends on their selection probability as shown in Figure~\ref{fig:mass_complete}, which we implement
as a linearly increasing probability with $p=0$ probability below the minimum detectable halo mass at each redshift, to $p=1$ at masses above the 100\% completeness 
threshold. In other words, for example, at low redshifts ($z<0.1$) even relatively low-mass groups with $\mhalo \sim 10^{13}\,\msun$ would not be counted in
\dmigm{}, since they can in principle be detected in our spectroscopy, and their \dmhalo{} contribution already taken into account. At higher redshifts, such low-mass 
groups are undetected and their influence would need to be considered as part of \dmaigm.
The effect of this halo clipping is to reduce both the mean \dmaigm{} and its variance. 

\begin{figure}
    \centering
    \includegraphics[width=0.5\textwidth]{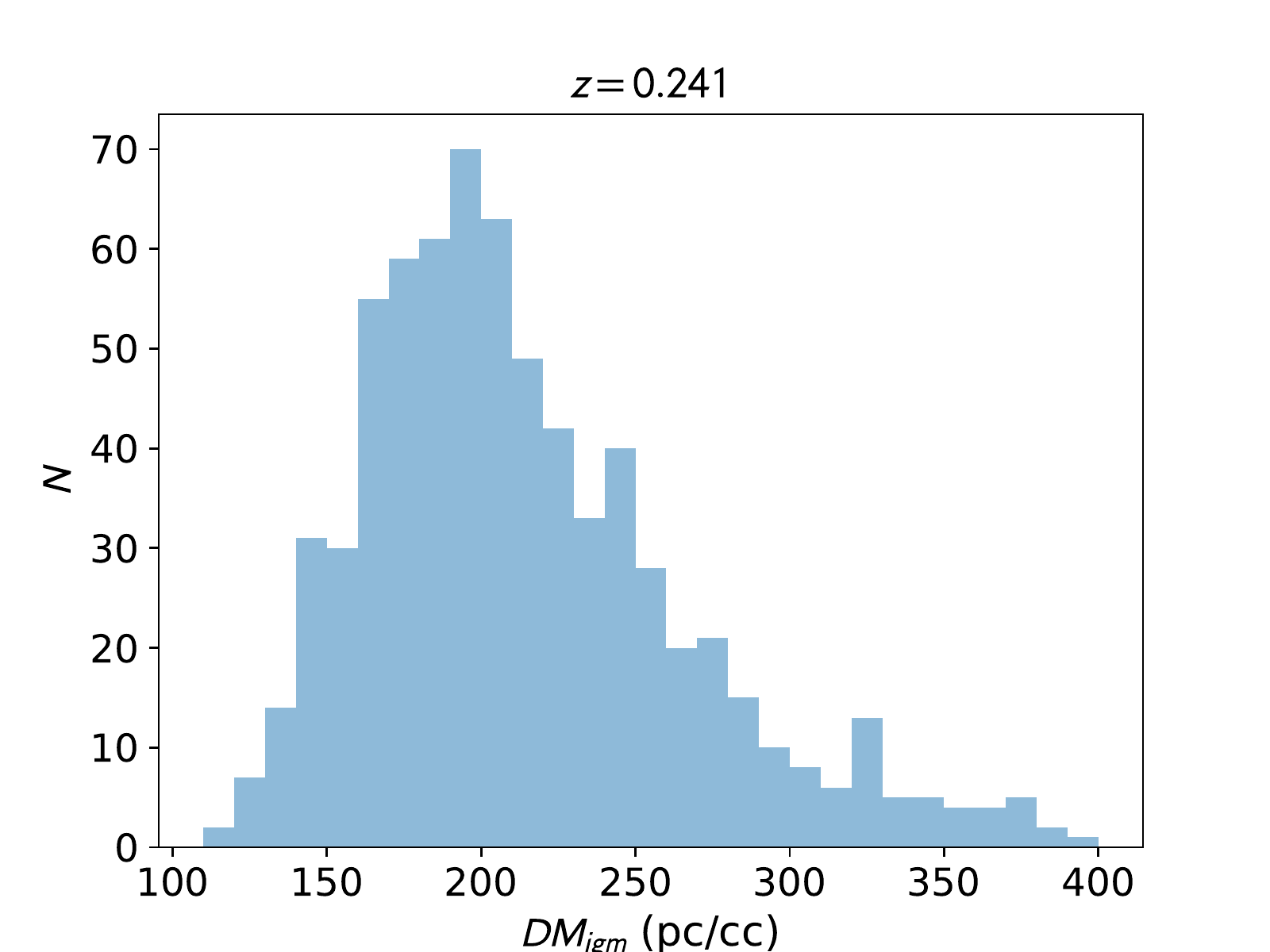}
    \caption{Distribution of \dmigm{} for $\zfrb=0.241$ sightlines in the Millennium simulation box, selected to intersect two $\mhalo > 10^{14}\,\msun$ halos. The direct influence of groups and halos 
    have been removed according to the selection function of Figure~\ref{fig:mass_complete}. We assume $\figm=0.85$ for this calculation.}
    \label{fig:dmigm}
\end{figure}

On the other hand, the \thisfrb{} sightline intersects two separate galaxy clusters, which means that it must be crossing through overdense regions of the Universe even if we have already removed the direct 
influence of the clusters themselves within $\sim 1-2\,\mathrm{Mpc}$.
For the \dmaigm{} calculation, we therefore selected mock sightlines at $\zfrb=0.241 \pm 0.001$ that 
intersect within \rvir{} of two clusters with $\mhalo > 10^{14}\,\msun$.
Compared with randomly-selected sightlines at this redshift selected to originate within galaxies with 
an observed magnitude of $r<20$ (see, e.g., \citealt{pol:2019}), 
we find 0.07\% of the sightlines intersect two separate foreground cluster halos, i.e.\ $\sim$ 1 in 1400.

We integrate the selected sightlines through Millenium density fields that have already been clipped 
of `observed' clusters, and compute \dmigm{} assuming $\figm=0.85$, which is a typical fraction of cosmic baryons expected to reside in the diffuse IGM as found in cosmological hydrodynamical simulations
\citep[e.g.,][]{jaroszynski:2019,batten:2021,takahashi:2021,zhu:2021,zhang:2021}.
The resulting \dmigm{} distribution is shown in Figure~\ref{fig:dmigm}, with a median $\dmaigm(z=0.241)=204^{+62}_{-39}\,\dmunits$, where errors are quoted at the 16th and 84th percentiles.
In comparison, for random sightlines at the same redshift through the normal `unclipped' density field we find $\dmaigm=191^{+64}_{-39}\,\dmunits$,
while for random sightlines through the clipped 
density fields we get $\dmaigm= 179^{+51}_{-33}$. The effects of clipping the influence of group/cluster
halos and selecting sightlines that go through two clusters thus appear to largely cancel each other out, although
the median \dmigm{} for our double-cluster sightlines is indeed slightly higher than the usual mean. 

\subsection{The Host DM of \thisfrb}
With the estimates of the foreground contributions in hand, we now subtract these from the total
dispersion measure of \thisfrb{} in order to constrain the host contribution. 
Since we the distributions for \dmigm{} and \dmhalos{} are significantly non-Gaussian, 
instead of propagating errors, we take the direct route of calculating the distribution of 
\dmhost{} based on the Monte Carlo realizations we have calculated for \dmigm{} and \dmhalos{}.

In other words, we compute multiple iterations of:

\begin{multline}\label{eq:dmhost}
    \mathrm{DM}_{\mathrm{host},i} = \mathrm{DM}_{\mathrm{FRB},i} -\mathrm{DM}_{\mathrm{MW},i}
- \mathrm{DM}_{\mathrm{IGM},i} \\ - \mathrm{DM}_{\mathrm{halos},i},
\end{multline}

\noindent where $\mathrm{DM}_{\mathrm{halos},i}$  and $\mathrm{DM}_{\mathrm{IGM},i}$ were drawn from the
Monte Carlo realizations computed in Sections~\ref{sec:halos} and \ref{sec:igm}, respectively.
For $\mathrm{DM}_{\mathrm{FRB},i}$ and $\mathrm{DM}_{\mathrm{MW},i}$, we draw Gaussian
random deviates for each iteration based on the values reported by \citet{niu:2022}: 
$\dmfrb=(1205\pm 4)\,\dmunits$ and $\dmmw=(113 \pm 17)\,\dmunits$, where the uncertainties
are adopted as the Gaussian standard deviations. 
We enforced the prior that $\dmhost\geq 0\,\dmunits$. 

After computing Equation~\ref{eq:dmhost} over $N=2000$ iterations, we computed the median and 16th/84th percentiles of the resulting $\mathrm{DM}_{\mathrm{host},i}$ distribution.
These quantities, as well as the individual DM components, are listed in Table~\ref{tab:dmhost}
--- note that since the medians are listed for each component, they do not necessarily obey Equation~\ref{eq:dm_all} precisely.

\begin{table}[ht]
\caption{\thisfrb{} DM Components \label{tab:dmhost}}
\setlength{\tabcolsep}{10pt}
\renewcommand{\arraystretch}{1.2}
\begin{tabular}{lcc}
Component\footnote{All DM values are in units of $\dmunits$} & \modelone & \modeltwo \\[3pt]
\hline\hline
\dmfrb{} & \multicolumn{2}{c}{$1205^{+4}_{-4}$} \\[3pt]
\hline
\dmmw{} & \multicolumn{2}{c}{$113^{+13}_{-13}$} \\[3pt]
\dmigm{} & \multicolumn{2}{c}{$204^{+62}_{-39}$} \\[3pt]
\dmhalos{} & $ 450^{+240}_{-130}$& $ 640^{+260}_{-150} $ \\[3pt]
\hline 
\dmhost{}\footnote{In the observed frame.} & $ 430^{+140}_{-220}$ & $ 280^{+140}_{-170}$
\end{tabular}

\end{table}

We separately calculated the \dmhost{} for the two different assumptions of \rmax{} for
the foreground halo contributions. 
Assuming $\rmax = \rvir$ (see Section~\ref{sec:halos}), we find $ \dmhost=430^{+140}_{-220}\,\dmunits$; on the other hand, for the $\rmax=2\,\rvir$ model, 
the resulting host contribution is $ \dmhost = 280^{+140}_{-170}\,\dmunits$ (both in the observed frame; the corresponding restframe values are $\dmhost= 540^{+170}_{-270}\,\dmunits$ and $ \dmhost=350^{+180}_{-210}\,\dmunits$, respectively).
These estimates are much lower than the value of $\dmhost =903^{+72}_{-111}\,\dmunits$ originally reported
by \citet{niu:2022}.
The 68th percentile errors we derive are significantly larger than the \citet{niu:2022} estimate, 
since they are driven by the uncertainties in the foreground cluster masses.
Adopting the upper 68th percentile errors of our \dmhost{} values as a $\sigma$, the original estimate for \dmhost{} is $3.4\sigma$ and
$4.3\sigma$ away from our \modelone{} and \modeltwo{} model estimates, respectively.

\section{Discussion}

The \dmhost{} of FRBs arise from all the free electron contributions in the host galaxy and the immediate vicinity of the FRB, starting from the so-far
mysterious source engine, ionized interstellar medium (ISM) gas, and then circumgalactic medium (CGM) of the host galaxy\footnote{If the FRB host galaxy is embedded in a galaxy cluster or group, then the cluster contribution is typically 
considered separately. See, e.g.\ \citet{connor:2023}.}. 
For localized FRBs with a clearly identified host galaxy, it is 
in principle straightforward to calculate the halo mass and estimate the halo CGM contribution \dmhostcgm{} the same way we did for the foreground halos, so we can write

\begin{equation}
    \dmhost=\dmhostcgm + \dmhostin,
\end{equation}

\noindent where we define \dmhostin{} as the `inner' DM components arising from the source engine and galaxy stellar/ISM component.

\subsection{Host Halo Contribution}\label{sec:hosthalo}
Given the association of \thisfrb{} with the $z=0.241$ host galaxy \frbhg{} by \citet{niu:2022}, we can start by estimating the contribution from the CGM of the host halo. 
For the reported stellar mass of $\mstar \approx 6\times10^8\,\msun$ \citep{niu:2022}, we
use the dwarf galaxy stellar mass-halo mass relationship of \citet{read:2017} to estimate a halo mass of $\mhalo = 9\times10^{10}\,\msun$. 
Allowing for the fact that the FRB originates 5kpc (\citealt{niu:2022}) from the galaxy center, and adopting the models
described in Section~\ref{sec:halos}, we estimate restframe CGM contributions of $\dmhostcgm=18\,\dmunits$ and $\dmhostcgm=12\,\dmunits$ assuming the \modelone{} and \modeltwo{}
models, respectively.
In other words, the CGM of \frbhg{} yields but a small contribution to
the \dmhost{} of \thisfrb{}. 
This allows us to estimate restframe values of the inner host contribution to be 
$\dmhostin \approx 520^{+170}_{-270}$ and $\dmhostin \approx 330^{+180}_{-210}$, 
where we have neglected the error on \dmhostcgm{} due to the small central value.

\subsection{Emission measures and scattering}\label{em_scat}
In their analysis of \thisfrb{}, \citet{niu:2022} and \citet{ocker:2022} had evaluated 
the emission measure (EM) from the observed optical H$\alpha$ lines
in \frbhg{},
which can be converted into an equivalent DM \citep{tendulkar:2017}.
This was found to yield an observed frame value in the range $\dmhost \approx 230-650\,\dmunits$, which was in tension with the original
estimate of $\dmhost=903\,\dmunits$. In comparison, {our new estimate for
the inner host contribution spans a 68 percentile confidence region of approximately $\dmhost \sim 110-690\,\dmunits$ (after combining results from both models in Section~\ref{sec:halos}).} The EM estimation is thus now
in agreement with \dmhost{} without having to invoke unusually high gas temperatures ($T \gg  10^4\,K$) in the H$\alpha$ emitting medium as suggested by \citet{ocker:2022}.
Given the broad agreement between the EM and DM estimates, the H$\alpha$ emitting gas --- presumably in the galaxy disk --- also now appears to make up the bulk of the ionized medium responsible for the dispersion seen in \frbhg{}. 

Across a large number of repeating signals, \thisfrb{} has also been measured by \citet{ocker:2022} to have a mean scattering
time delay of $\tau=10.9\pm1.5\,\ms$ (measured at 1.41 GHz) that can be attributed to the host. 
This DM of the scattering screen can be estimated from the scattering timescale using the expression from \citet{cordes:2022}:

\begin{equation}
\tau(\mathrm{DM}, \nu, z) \approx 48.03\,\mu s \;\times\; \frac{A_\tau \effgee\, \mathrm{DM}^2}{(1+\zfrb)^3\,\nu^4},
\end{equation}

\noindent where $\nu$ is the measured frequency in GHz. The $A_\tau$ is a dimensionless quantity relating the mean scattering delay to the
$1/e$ time, which we assume to be $A_\tau \approx 1$ following \citet{ocker:2022}. 
The combined factor \effgee{} describes the combined amplification from turbulent density fluctuations and
geometry of the scattering screen, respectively, which \citet{ocker:2022} estimate to be $\effgee = 1.5^{+0.8}_{-0.3}\,(\mathrm{pc^2\;km})^{-1/3}$
using the old value of $\dmhost = 903\,\dmunits$. 

With our updated restframe \dmhostin{} values for \thisfrb{} (Section~\ref{sec:hosthalo}), we recalculate $\effgee \propto 1/\mathrm{DM}^2$ to find $\effgee \approx 4.5 - 11\,(\mathrm{pc^2\;km})^{-1/3}$, 
with the range allowing for our model uncertainty in subtracting off the foreground galaxy clusters. This constrasts with the value originally estimated by \citet{ocker:2022}, which is close to unity. 
This larger value implies that either $\eff>1$ or $G>1$, or both. 
If $\eff > 1$ it would imply that the scattering screen is highly turbulent. Meanwhile, $G\approx 1$ is expected if the turbulent scattering screen is close to the source, but could be greater than unity if the screen is somewhere along the intervening path yet still farther away from the Milky Way, e.g.\ if they were associated with the foreground clusters.
However, \citet{connor:2023} recently studied two FRBs that were
localized to host galaxies embedded within galaxy clusters, and did not see significant scattering in those FRB signals. 
We therefore consider it unlikely that foreground clusters are the source of the scattering; it is more likely that the large $\effgee$ value is due to a highly turbulent scattering screen associated with the host and perhaps even close to the FRB engine itself.
{
This conclusion is corroborated by the recent \citet{ocker:2023} paper which reports significant scattering variations in the repeated \thisfrb{} signals, as well as 
the sign-changes observed in their Faraday rotation \citep{anna-thomas:2023}.
}

% %%%%%%%%%%%%%%%%%%%%%%%%%%%%%%%%%%%%%%%%%%%
\subsection{\thisfrb{} in Context}
When \thisfrb{} was discovered, its \dmhost{} estimate and those of nearly all other localized FRBs were done through guesstimated 
\dmhost{} values after subtracting the mean $\dmaigm(z)$. 
At this time of writing, approximately a half-dozen localized FRBs now have credible foreground analyses that enable more confidence in their estimated \dmhost{}. 

Arguably the first reliable estimate was done by \citet{simha:2020}, who analyzed the foreground of FRB20190608 based on Sloan Digital Sky Survey (SDSS) and Keck data using an analogous technique to ours. They estimate $\dmhost \approx 80-200\,\dmunits$ (observed frame), which is in fact consistent with an independent analysis of the host galaxy \citep{chittidi:2021}.

More recently, \citet{simha:2023} analyzed 4 FRB sightlines known to
exhibit DMs significantly higher than the cosmic mean, using FLIMFLAM
spectroscopic data of the foreground fields. 

FRB20210117A was, like \thisfrb{}, suspected to be a high \dmhost{} source given its clear excess $\dmfrb= 731\,\dmunits$ and its localized redshift of $\zfrb = 0.2145$. 
However, unlike \thisfrb{} no significant excess was found from \dmhalos{} based on the foreground data, and so it is confirmed to be a high-\dmhost{} source, with $\dmhost \approx 665\,\dmunits$ in the restframe. 
This is in fact higher than the revised restframe values of $\dmhost \sim 350-540\,\dmunits$ we now find for \thisfrb{}, which now makes FRB20210117A in principle the
FRB with the highest known \dmhost{}, although the uncertainties are large enough for either to be the true record holder. 
FRB20200906A was also shown to not have significant foreground contributions despite a high \dmfrb{}, yielding an estimate of $\dmhost \approx 420\,\dmunits$.
FRB20190714A and FRB20200430A are shown to have significant foreground contributions, thus they do not have large \dmhost{}. 
We note that this sub-sample is explicitly biased towards excess-DM sources, and thus it possible that the \dmhost{} from this sample might be biased high as well even though some of the FRBs are shown to be from overdense foregrounds.

\cite{james:2022b} and subsequently
\cite{baptista:2023} % https://ui.adsabs.harvard.edu/abs/2023arXiv230507022B/exportcitation
have performed population modeling
of a sample of $\approx 70$~FRBs including 21 with redshifts
from host associations.  Their forward model includes
two parameters to desribe a log-normal distribution
for \dmhost.
Taking their preferred values of $\sigma = 0.5$
and $\mu = 2.43$, we find that 40\%\ of FRB hosts
are expected to have
$\dmhost > 500 \, \dmunits$.
We conclude, therefore, that our inferred value for
FRB20190520B is consistent with the full
population.  
%We further note that the CHIME collaboration
%has reached a similar conclusion for the population \citep{masoud}. % https://ui.adsabs.harvard.edu/abs/2021ApJ...922...42R/exportcitation

\section{Conclusion}
In this paper, we used wide-field spectroscopic data from the FLIMFLAM survey targeting the field of \thisfrb{} to study the
possible foreground contributions to the overall observed DM, which is anomalously high (\dmfrb = 1205\,\dmunits)
given its confirmed redshift.
Our data shows that the \thisfrb{} sightline directly intersects two foreground galaxy clusters, at $z=0.1867$ and $z=0.2170$,
within their virial radius, in a rare occurrence estimated to happen to only $\sim$1 in 1400 sightlines
at this FRB pathlength.
These two foreground clusters yield a combined contribution of $\dmhalos \sim 450-640\,\dmunits$, 
which allows us to revise the host contribution downwards to $\dmhost \sim 280-430\,\dmunits$ after also taking account the
Milky Way and diffuse IGM contributions (see Table~\ref{tab:dmhost} for detailed values). 
This means that \thisfrb{} is no longer the FRB with the largest host DM contribution, with FRB20210117A now being the source with the largest known \dmhost{}, but they are similar to within the uncertainties. 
The new value for \dmhost{} is now in agreement with H$\alpha$ emission estimates of \dmhost{}, and also allowed us to make a revised 
estimate of the combined factor describing geometric effects and turbulent density fluctuations based on the scattering time scale: 
$\effgee \approx 4.5 - 11\,(\mathrm{pc^2\;km})^{-1/3}$.
We interpret this as due to high turbulence in the scattering screen associated with the host, since we consider it unlikely that the foreground halos are the source of the scattering.

This result outlines the importance of obtaining sufficiently wide-field foreground spectroscopy in disentangling the various DM contributions in FRBs, 
both for their usage as cosmological probes as well as for understanding their host progenitors. 
The original Keck/LRIS optical spectroscopy from \citet{niu:2022} were single-slit observations, but
even if they had utilized multi-object slitmasks to target some of the foreground galaxies, they would
have been limited to within 2-3$\arcmin$ of the FRB position by the limited field of view of LRIS
(see the interactive version of Figure~\ref{fig:field_bokeh}). LRIS observations might have captured 2-3 of the $\zgrp=0.1867$ group members, but the rest lie 4-5$\arcmin$ away from the FRB. The member galaxies of the $\zgrp=0.2170$ group, on the other hand, lie mostly $\sim 5-6\arcmin$ away from \thisfrb{} and would likely have been
missed even with multi-object spectroscopy with LRIS or equivalent
narrow-field spectrographs on other telescopes. 

Our present result on \thisfrb{} is a vivid illustration of the power of foreground data in enhancing the use of FRBs.
In \citet{lee:2022}, we had estimated that wide-field foreground spectroscopy enhances the power of localized FRBs as cosmological probes by a factor of $\sim30$
in terms of the number of FRBs required to achieve comparable constraints ---
whereas in comparison with \emph{unlocalized} FRBs, this is an enhancement is up to $\sim 10^2-10^3$ \citep{shirasaki:2022, wu:2023}.

In an upcoming paper (Khrykin et al, in prep), we will present the first cosmic baryon analysis from a preliminary sample of FLIMFLAM fields, that will
give the first constraints on the partition of cosmic baryons between the IGM and CGM.

\section*{Acknowledgements} 
\label{sec:acknowledgements}
We are grateful to Elmo Tempel for kindly sharing his group finding code, to Chris Lidman for helping with the data reduction code, and to various members of the CRAFT collaboration for useful discussions.
K.N. acknowledges support from MEXT/JSPS KAKENHI grants JP17H01111, 19H05810, and 20H00180, as well as travel support from Kavli IPMU, World Premier Research Center Initiative, where part of this work was conducted.
We acknowledge generous financial support from Kavli IPMU that made FLIMFLAM possible. Kavli IPMU is supported by World Premier International Research Center Initiative (WPI), MEXT, Japan. Based on data acquired at the Anglo-Australian Telescope, under programs A/2020B/04, A/2021A/13, and O/2021A/3001. We acknowledge the traditional custodians of the land on which the AAT stands, the Gamilaraay people, and pay our respects to elders past and present.
Authors S.S., J.X.P., I.K., and N.T., 
as members of the Fast and Fortunate for FRB
Follow-up team, acknowledge support from 
NSF grants AST-1911140, AST-1910471
and AST-2206490. 

\software{
configure \citep{miszalski:2006},
MARZ \citep{hinton:2016},
CIGALE \citep{boquien:2019},
Astropy \citep{astropy-collaboration:2022},
Numpy \citep{harris:2020},
Scipy \citep{virtanen:2020},
Matplotlib \citep{hunter:2007},
Bokeh.
}

\bibliography{references_kg}
\end{document}